\documentclass[preprint,showpacs,aps,prd,nofootinbib,floatfix,amsmath,amssymb]{revtex4}

\usepackage{graphicx}
\usepackage{color}
\usepackage{subfig}
\usepackage{multirow}
\usepackage{slashed}
\usepackage{color}
\usepackage{hyperref}
\newcommand{\splash}[1]{{#1\mkern-9.0mu /}}
\usepackage{cancel}

\makeatletter
\newbox\slashbox \setbox\slashbox=\hbox{$/$}
\newbox\Slashbox \setbox\Slashbox=\hbox{\large$/$}
\def\pFMslash#1{\setbox\@tempboxa=\hbox{$#1$}
	\@tempdima=0.5\wd\slashbox \advance\@tempdima 0.5\wd\@tempboxa
	\copy\slashbox \kern-\@tempdima \box\@tempboxa}
\def\pFMSlash#1{\setbox\@tempboxa=\hbox{$#1$}
	\@tempdima=0.5\wd\Slashbox \advance\@tempdima 0.5\wd\@tempboxa
	\copy\Slashbox \kern-\@tempdima \box\@tempboxa}

\def\miss#1{\ifmmode{/\mkern-11mu #1}\else{${/\mkern-11mu #1}$}\fi}
\makeatother

\begin{document}
	
\title{Dipolar electroweak properties of Dirac massive neutrinos}

\author{J. P. Guti\'errez-Montes$^{1}$, J. Monta\~no-Dom\'inguez$^{1,2}$,  F. Ram\'irez-Zavaleta$^1$,  E. S. Tututi$^1$ and  O. V\'azquez-Hern\'andez$^{1,2}$}

\address{$^1$Facultad de Ciencias F\'{i}sico Matem\'{a}ticas,
	Universidad Michoacana de San Nicol\'{a}s de Hidalgo, Av. Francisco J. M\'{u}jica S/N, 58060, Morelia, Michoac\'{a}n, M\'{e}xico, }

\address{$^2$SECIHTI, Av. Insurgentes Sur 1582, Col. Cr\'edito Constructor,
Alc. Benito Ju\'arez, C. P. 03940, CDMX, M\'exico.
\\email: 0806477x@umich.mx,
 jmontano@secihti.mx, iguazu.ramirez@umich.mx, oscar.vazquez@umich.mx, eduardo.tututi@umich.mx}

\date{August, 2025}
\begin{abstract}
 Electroweak properties of Dirac neutrinos within the context of the Minimally Extended Standard Model are explored. In particular, the dipolar form factors  that result from  the radiative correction, at one-loop level, of the $ Z\nu \bar{\nu}$ coupling are identified. The calculation is carried out in the covariant general gauge $R_{\xi}$. The results are shown to be independent of the gauge parameter. 
The outcome   numerical evaluation establishes that the neutrino weak-magnetic dipole moment is  of the same order of magnitude as the known prediction of the neutrino magnetic dipole moment. 

\end{abstract}	
	
\maketitle

\section{INTRODUCTION}

Transitions that violate lepton flavor, such as $\mu\to e \gamma$,  arise at  the one-loop level   in   theories beyond the Standard Model (SM), where  it is demonstrated that these interactions  can be   understood  as  electromagnetic dipolar  transitions \cite{cheng-li,marciano-sanda,petcov,lee-shrock,novales}.  These processes   are characterized by dimension-five operators  of the form $\bar{\nu}\sigma^{\mu \nu}(F_M+F_E\gamma^5 )\nu\,F_{\mu \nu}$, being $F_M$ and $F_E$  the magnetic and electric form factors, respectively, and  $F^{\mu\nu}$ is  the electromagnetic  field tensor.  Similar Lorentz structures were also used to study electromagnetic transitions between massive neutrinos $\nu_i\to\nu_j\gamma$ \cite{sato-kobayashi,goldman-stephenson,pal-wolfenstein}  ($i,j=1,2,3$) along with  its magnetic  dipole  moments  (for the diagonal case $i=j$) \cite{marciano-sanda,petcov,shrock,Fujikawa:1980yx,Giunti:2014ixa,omar}.  In simple extensions to the  SM  the value of the  electromagnetic
dipole moment (EMDM) for Dirac neutrinos results very small in comparison with   the Bohr magneton $\mu_B$:  $\mu_{\nu_{i}}\approx \frac{3eG_Fm_{\nu_i}}{8\sqrt{2}\pi^2}\approx 3.2\times 10^{-19}\left(\frac{m_{\nu_i}}{\mathrm{1 eV}}\right)\mu_B$, where $m_{\nu_i}$ is the mass of the neutrino and, as usual, $G_F$ is the Fermi constant.  On the contrary, because  the Majorana neutrino is its own anti-particle, the CPT symmetry  demands that its electromagnetic dipole moments  are zero. There are different contexts where the  EMDM of neutrinos have also  been studied to constrain its numerical value.  For example, in astrophysical studies, the plasmon decay into pair of neutrinos, $\gamma^*\to\nu\bar{\nu}$, which is responsible for energy loss in stars \cite{bernstein,raffelt}, can be understood in terms of the neutrino dipolar magnetic moment. In cosmological contexts, from high precision cosmic microwave background spectral data, it was established  an upper bound  for an effective 
neutrino magnetic moment of the order of $10^{-8}\mu_B$ \cite{mirizzi}.  In extra dimensions \cite{mohapatra}, it is calculated   the  $\overline{\nu}_e e\to e \overline{N}$ cross section (where $N$  is a right-handed massive neutrino) taking into account the dipolar moment of the neutrino to constraint on a class of large extra dimensions.

On the other hand, weak properties of charged fermions such as the weak dipole moments, which results from quantum fluctuations and contribute to the $Zf\bar{f}$  (with $f$ denoting any quark or charged lepton) coupling, have been studied in different context beyond the SM.
Due to the symmetry properties of weak interactions,  the  resulting Lorentz structure is  similar to the corresponding electromagnetic Lorentz structure. In fact,  analogous  weak form factors,  which are found  by replacing  $F^{\mu\nu}$ by the corresponding weak field tensor $Z^{\mu\nu}=\partial^{\mu}Z^{\nu}-\partial^{\nu}Z^{\mu}$, being $Z^\mu$ the weak field of the $Z$ gauge boson, have  also been used to calculate  weak magnetic dipole moments (WMDM)  of charged fermions \cite{Bernabeu:1995gs,Bernabeu:1994wh,Bernabeu:1997je,Hollik:1997vb,deCarlos:1997br,Hollik:1997ph,GomezDumm:1999tz,Bolanos:2013tda,Arroyo-Urena:2016ygo,Arroyo-Urena:2017sfb,Arroyo-Urena:2015uoa,Hollik:1998vz,Moyotl:2012zz,montano:2022}.  
For the case of charged leptons, it is interesting the anomalous weak dipole moments of the tau lepton for which there are experimental bounds; these are    such that Re$(a_\tau^{w}) < 1.1 \times 10^{-3}$ at 95\% C.L., and Im$(a_\tau^{w}) <2.7\times 10^{-3}$ at 95 \% C.L. In contrast,  the value for the  WEDM is given by  Re$(d_\tau^{w}) < 0.5\times10^{-17} e$$\cdot$cm at 95\% C.L., and Im$(d_\tau^{w}) < 1.1 \times ^{-17} e$$\cdot$cm at 95\% C.L. \cite{aleph-2003,PDG}.

Thus,  the weak properties of neutrinos such as the   weak dipole moments and its possible consequences in  experimental, cosmological, astrophysical and others contexts result intriguing  and deserve attention.
However, as far as the available literature is concerned on the subject, the WMDM of neutrinos have not been yet analyzed.
In this work we study the weak dipole moment of Dirac neutrinos which is carried out within the framework of the Minimally Extended Standard Model (MESM) \cite{Giunti-Kim} that includes right-hand components of the chiral fields of neutrinos and thereby allows mass terms for neutrinos. The calculation of the   neutrino WMDM is carried out  at the one-loop level in a general gauge and it is shown that the neutrino weak dipole moment is ultraviolet finite and gauge independent.

\section{Basic formalism}
\subsection{The Minimally Extend Standard Model in a $R_\xi$ gauge}
Let us briefly discuss the MESM in which it is included three right-handed neutrinos into the SM  \cite{cheng-li,Giunti-Kim}. In order to analyze the gauge independence of the weak dipole moments of neutrinos, we consider that the corresponding gauge sector  is furnished with a general  $R_\xi$ gauge by means of the   $\xi$ parameter. In particular, we focus on the Yukawa sector, since other sectors of the MESM turns out be the same as the SM in the $R_\xi$ gauge (${\cal L}^{MESM}_{R_\xi}$), which is discussed, for example,  in Refs. \cite{cheng-li,fujikawa,toscano-2004}. For this case, the Yukawa Lagrangian  takes the form
\begin{equation}
\label{YL}
{\cal L}^Y_{R_\xi}=-\sum_{\alpha,\beta=e,\mu,\tau}Y_{\alpha\beta}^{\prime l}\overline{L_{L\alpha}^{\prime}}\,\Phi \,l_{R \beta }^{\prime}
-\sum_{\alpha,\beta=e,\mu\tau}Y_{\alpha\beta}^{\prime \nu}\overline{L_{L\alpha}^{\prime}}\,\tilde{\Phi} \,\nu_{R \beta }^{\prime}+\mathrm{H.c.} \subset {\cal L}^{MESM}_{R_\xi},
\end{equation}
where 
\[
L_{L\alpha}^{\prime}=\left(\begin{array}{c} \nu_{L\alpha}^{\prime}\\ \l_{L\alpha}^{\prime}\end{array}\right),\qquad
\Phi=\left(\begin{array}{c} \phi^+\\ \frac{\phi_1+i\phi_2}{\sqrt{2}}\end{array}\right),
\]
 are left-handed $SU(2)$ weak doublets and the Higgs doublet with hypercharge $Y=1$, respectively.
The prime denotes Dirac fields in the gauge representation, the various  $Y_{\alpha\beta}^{\prime l (\nu)}$ represent the elements of $3\times 3$ matrices of Yukawa couplings for leptons (neutrinos), $ l_{L(R),\alpha}^{\prime}=e_{L(R)}^{\prime},\mu_{L(R)}^{\prime},\tau_{L(R)}^{\prime}$ are the left (right) components of the  lepton fields, $ \nu_{L(R),\alpha}^{\prime}=\nu_{e\,L(R)}^{\prime},\nu_{\mu\,L(R)}^{\prime},\nu_{\tau\,L(R)}^{\prime}$  are the respective left (right) components of the neutrinos in the gauge representation and $\tilde{\Phi}=i\tau_2\Phi$, being $\tau_2$ the second Pauli matrix. After the spontaneous symmetry breaking, the Higgs doublet can be written as
\begin{equation}
\label{asbs}
\Phi=\left(\begin{array}{c} G_W^+\\ \frac{v+H+iG_Z}{\sqrt{2}}\end{array}\right),
\end{equation}
where $v$ is the vacuum  expectation value (VEV) of the Higgs doublet;  $G_W^+, G_Z$ and $H$ are the charged  and neutral unphysical Goldstone  bosons and the Higgs field, respectively. Let us now move to the representation where the leptons have definite mass by means of a bi-unitary transformation by diagonalizing the Yukawa couplings:
\begin{eqnarray}
\label{bi-unitary }
Y^{\prime l }\to Y_D^{l}&=&U_L^{l \dag}Y^{\prime l }U_R^{l },\qquad [Y^{ l }_D]_{\alpha\beta}=y^{l}_{\alpha}\delta_{\alpha\beta},\nonumber\\
Y^{\prime\nu}\to Y_D^{\nu}&=&U_L^{\nu\dag}Y^{\prime \nu}U_R^{\nu},\qquad [Y^{\nu}_D]_{ij}=y^{\nu}_{i}\delta_{ij},\quad  i,j=1,2,3.
\end{eqnarray} 
In last equation, $U_{L,R}^{l}\,(U_{L,R}^{\nu})$ represent appropriate unitary matrices that diagonalize the matrices $Y^{\prime l }\, (Y^{\prime \nu})$ for charged leptons (neutrinos). Thus, the diagonalized Yukawa Lagrangian for leptons can be expressed as
\begin{eqnarray}
\label{YLASSB}
{\cal L}^Y_{R_\xi}&=&-\frac{1}{\sqrt{2}}\left(v+H+iG_Z\right)\left[\sum_{\alpha=e,\mu\tau}y_{\alpha}^l\overline{l_{L\alpha}}\,l_{R \alpha }
+\sum_{i=1,2,3}y_{i}^{\nu}\overline{\nu_{L i}}\,\nu_{R i }\right],\nonumber\\
&+&G_W^{+}\sum_{\alpha, i}\left(y_{\alpha}^l[U]_{\alpha\,i}\overline{l_{L\alpha}}\,\nu_{R i }-y_{i}^{\nu}[U^{\dag}]_{\alpha\,i}\overline{\nu_{L i}}\,\nu_{R i }\right) +\mathrm{H.c.},
\end{eqnarray}
where
\begin{equation}
\label{transfo}
l_{L(R) \alpha}=[U^{l \dag}_{L(R)}]_{\alpha\beta}l_{L(R) \beta}^{\prime},\qquad  \nu_{L(R) i}=[U^{\nu \dag}_{L(R)}]_{i \alpha}\nu_{L(R) \alpha}^{\prime},
\end{equation}
here, a sum over repeated indexes is implied and the lepton fields $l_{L(R) \alpha}$ and $\nu_{L(R) i}$ are in the representation of definite mass. The matrix  $[U]_{\alpha\,i}\equiv [U_{L}^{l \dag}]_{\alpha \beta}[U_{L}^{\nu}]_{\beta i}$ in Eq. (\ref{YLASSB})  is the lepton matrix mixing analogous to the CKM matrix  of quarks. From Eq. (\ref{YLASSB}), it can be obtained directly the Feynman rules for the couplings between the Higgs, Goldstone bosons and the leptons.  The  corresponding lepton  masses are defined by considering that, for example, for charged leptons $m_\alpha=y^l_\alpha v/\sqrt{2}$ and  similarly for the neutrino masses: $m_i=y^\nu_i v/\sqrt{2}$. As usual, the VEV is related to the $W$ gauge boson as: $m_W=g v/2$, being $g$ the weak coupling constant.

\subsection{An effective Hamiltonian}

Let us  consider the effective Hamiltonian that accounts massive Dirac neutrinos interacting  weakly   through  a  $Z$ gauge boson in the MESM. For this case, the Hamiltonian can be written as
\begin{equation}
\label{heffective }
{\cal H}^{(\nu)}(x)=-{\cal L}^{(\nu)}(x)=J^{(\nu)}_\alpha(x)Z^{\alpha}(x)=\overline{\nu}_i(x)\Gamma_{\alpha}^{ij}\nu_j(x)Z^{\alpha}(x).
\end{equation}
 Here, a sum over repeated index is implied ($i,j=1,2,3$), $J^{(\nu)}_\alpha(x)$ is the neutrino effective weak four-vector current and $\Gamma_{\alpha}^{ij}$ is a $4\times 4$ matrix in spinor space, that in general depends on space-time derivatives, such that it transforms as a four-vector. In space moment, the most general form factors in the Lorentz decomposition   of $\Gamma_{\alpha}^{ij}$, taking into account the hermicity  and current conservation properties,   can be conveniently expressed as 
\begin{equation}
\label{vertex1}
\Gamma_{\mu}^{ij}=\left[F_Q^{ij}(q^2)+F_A^{ij}(q^{2})\gamma_5\right]\left(q^2\gamma_\mu-\splash{q} q_{\mu}\right)-\left[F_M^{ij}(q^2)+iF_E^{ij}(q^2)\gamma_5\right]i\sigma_{\mu\nu}q^\nu,
\end{equation}
where the various form factors $F_X^{ij}(q^2)$  ($X=Q,A,E,M$) are  $3\times3$ Hermitian matrices    in the flavor space and the low-indices identify the   different weak properties to the neutrino, being $q$ the transferred moment.  For the diagonal elements, with $i=j$, the  form factors represent:  the charge, the anapole, the electric and magnetic contributions, respectively. Notice that the vertex $\Gamma_{\alpha}^{ij}$ is quit similar to the respective neutrino electromagnetic vertex, however, now the corresponding form factors are for the  weak interaction. For our interest, we drop the first term in Eq. (\ref{vertex1}), which is analogous of the anapole moment in electromagnetic interactions, and concentrate in the second term  by defining an effective Lagrangian with it.

\section{The anomalous weak magnetic dipole moment}

According  to Eq. (\ref{vertex1}), we can define  the five-dimensional  effective Lagrangian \cite{Roberts-marciano}, consistent with Lorentz and electroweak gauge invariance, that characterizes the weak dipolar interaction $Z\nu_i\bar{\nu_i}$ 
\begin{equation}\label{}
\mathcal{L}=-\frac{1}{2}\bar{\nu}_i\sigma^{\mu\nu}(\mu_{\nu i}^w-id_{\nu i}^w\gamma^5)\nu_i Z_{\mu\nu}, 
\end{equation}
where $\mu_{\nu i}^w=2F_M^{ii}(m_W^2)$  and $d_{\nu i}^w=-2F_E^{ii}(m_W^2)$  are the anomalous weak magnetic (AWMDM) and weak electric dipole moments (AWEDM), respectively, which  have units of inverse mass. The dimensionless AWMDM,  $a_{\nu i}^w$,  can be obtained via  the   dipole moment $\mu_{\nu i}^w=e a_{\nu i}^w/(2m_{\nu i})$.
The corresponding on-shell vertex function or Lorentz structure, illustrated in Fig.~\ref{FIGURE-vertex-1}, is
\begin{equation}\label{}
\Gamma^\mu=\sigma^{\mu\nu}q_\nu(\mu_{\nu i}^w-id_{\nu i}^w\gamma^5).
\end{equation}
\begin{center}
\renewcommand{\thesubfigure}{\arabic{subfigure}}
\begin{figure}[!ht]
\includegraphics[scale=0.7]{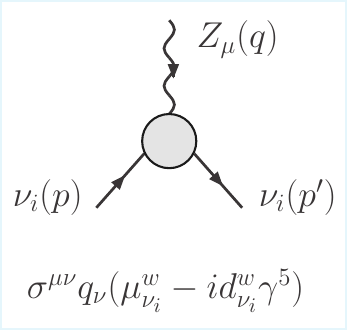}   
\caption{Feynman rule of the dipolar interaction $Z\nu\bar{\nu}$, where $p+q=p'$.}
\label{FIGURE-vertex-1}
\end{figure}
\end{center}
At the one-loop level, the vertex function  that gives rise to the AWMDM of the Dirac neutrino, $\mu_{\nu i}^w$, receives contributions from four sets of loop diagrams with virtual bosons, characterized  by $Z$, $W$, $Z$-$H$, and $H$; they are depicted in the Fig.~\ref{FIGURE-loops} in the covariant $R_\xi$ gauge. In advance, $\mu_{\nu i}^w$ will results independent of the gauge parameter $\xi$, and the expected absence of $d_{\nu i}^w=0$. We have cross-checked with the unitary gauge, $\xi\to\infty$, obtaining the same result. The AWMDM can be expressed as
\begin{eqnarray}\label{AWMDM}
\mu_{\nu i}^w &=& \mu_{\nu i}^w(Z)+\mu_{\nu i}^w(W)+\mu_{\nu i}^w(Z\text{-}H)+\mu_{\nu i}^w(H).
\end{eqnarray}
We must keep in mind that a consequence of introducing right-handed neutrinos in the SM implies the appearing of the $H\nu\bar{\nu}$ interaction, in the same way as the rest of the $Hf\bar{f}$ couplings. Thus, $H\nu\bar{\nu}$ is responsible  of the $\mu_{\nu i}^w(Z\text{-}H)$ and $\mu_{\nu i}^w(H)$ contributions. In the following, we detail analytically the leading contribution of each set of diagrams, where we have taken into account that  $m_{\nu i,l_k}\ll m_{W,Z,H}$.
\begin{center}
\renewcommand{\thesubfigure}{\arabic{subfigure}}
\begin{figure}[!t]
{\includegraphics[scale=1]{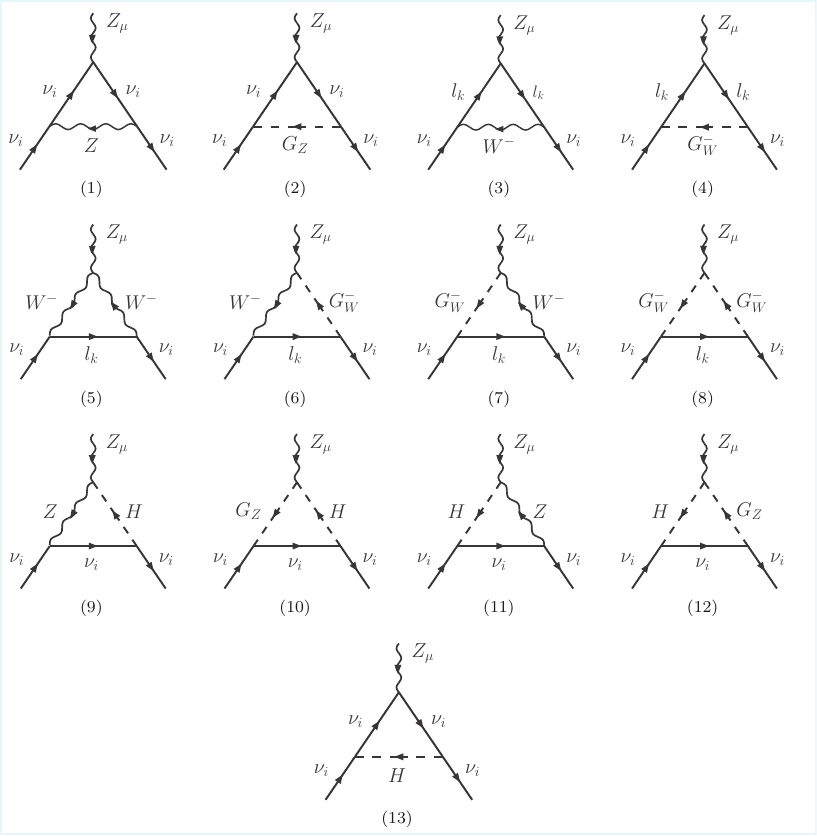}}  
\caption{1-loop contributions to the Dirac massive neutrino weak dipole moment in the covariant $R_\xi$ gauge.}
\label{FIGURE-loops}
\end{figure}
\end{center}

The set of the $Z$ virtual contribution and its pseudo-Goldstone boson $G_Z$, Fig.~\ref{FIGURE-loops} (1)-(2), results in
\begin{eqnarray}\label{Z-contribution}
\mu_{\nu i}^w(Z) &=& -\frac{eG_F}{96\sqrt{2}\pi^2c_Ws_W}m_{\nu i}
\left(-75+32\pi^2-96\text{Li}_22-66 i\pi\right).
\end{eqnarray}
Here, the $\mathrm{Li}_2(x)$ is the dilogarithmic function, $c_W$ and $s_W$  stand for $\cos\theta_W$ and $\sin\theta_W$, being $\theta_W$  the Weinberg angle.
The set of the $W$ contribution and its pseudo-Goldstone boson $G_W$, Fig.~\ref{FIGURE-loops} (3)-(8), yields
\begin{eqnarray}\label{W-contribution}
\mu_{\nu i}^w(W) &=& \frac{eG_Fc_W}{48\sqrt{2}\pi^2s_W}m_{\nu i} 
\big[
72c_W^4+c_W^2\left(72s_W^2-6\right)+78s_W^2-51
+\left(2s_W^2-1\right)
\Big\{6i\pi\left(6c_W^2+5\right)
\nonumber \\
&& +\left(72c_W^2+60\right)\ln c_W
+\left(3c_W^4+4 c_W^2+1\right)\left[12\text{Li}_2\left(c_W^2+1\right)+24\ln^2c_W+24i\pi\ln c_W-4\pi ^2\right]
\Big\}
\nonumber\\
&&
+12\left(6c_W^4+3c_W^2-1\right)R_W
\ln\frac{2 c_W^2-1+R_W}{2c_W^2}
-12\left(6c_W^4+2c_W^2-1\right)
m_W^2C_0^W
\bigg],
\end{eqnarray}
where $R_W\equiv \sqrt{1-4c_W^2}$, and the three-point Passarino-Veltman scalar function given in the notation of \texttt{Package-X} \cite{packageX}
\begin{eqnarray}
m_W^2C_0^W &=& m_W^2C_0(0,0,m_Z^2;m_W,0,m_W)
\nonumber\\
&=& c_W^2\bigg(-\frac{\pi^2}{6}+\text{Li}_2\frac{2 c_W^2}{2 c_W^2+R_W-1}
+\text{Li}_2\frac{2-2 c_W^2}{R_W+1}
-\text{Li}_2\frac{2-2 c_W^2}{-2 c_W^2+R_W+1}
+\text{Li}_2\frac{c_W^2-1}{c_W^2}
\nonumber\\
&& +\frac{1}{2}\ln^2\frac{2 c_W^2+R_W-1}{R_W+1}
-\text{Li}_2\frac{1-R_W}{2}
-\frac{1}{2}\ln^2\frac{R_W+1}{2}\bigg).
\end{eqnarray}
The set of the virtual combination $Z$-$H$ and the corresponding $G_Z$-$H$, Fig.~\ref{FIGURE-loops} (9)-(12), is
\begin{eqnarray}\label{Z-H-contribution}
\mu_{\nu i}^w(Z\text{-}H) &=&
\frac{eG_F}{8 \sqrt{2} \pi^2 c_W s_W}m_{\nu i}
\bigg(
-1
+r_H^2\ln r_H-iR_H\thinspace r_H \ln\frac{r_H+iR_H}{2}
+r_H^2m_Z^2C_0^{ZH}
\bigg),
\end{eqnarray}
where $r_H\equiv m_H/m_Z$, $R_H\equiv\sqrt{4-r_H^2}$,
\begin{eqnarray}
m_H^2 C_0^{ZH} &=& m_H^2C_0(0,0,m_Z^2;m_Z,0,m_H)
\nonumber\\
&=&
-\frac{\pi ^2}{6}
-\text{Li}_2\frac{2\left(r_H^2-1\right)}{r_H\left(r_H-iR_H\right)}
+\text{Li}_2\frac{r_H^2-1}{r_H^2}
+\text{Li}_2\frac{2r_H}{r_H+i R_H}
\nonumber\\
&&
-\text{Li}_2\frac{2\left(r_H^2-1\right)}{r_H\left(r_H+iR_H\right)}
+\text{Li}_2\frac{2r_H}{r_H-iR_H}.
\end{eqnarray}
The pure $H$ virtual contribution, Fig.~\ref{FIGURE-loops} (13), provides
\begin{eqnarray}\label{H-contribution}
\mu_{\nu i}^w(H) &=& -\frac{eG_F}{64\sqrt{2}\pi^2c_Ws_W}\frac{m_{\nu i}^3}{m_Z^2}
\bigg\{
4 r_H^2\left(r_H^2+1\right)\left[\pi^2-3 \text{Li}_2\left(r_H^2+1\right)-6\ln^2r_H\right]
-9-12r_H^2
\nonumber\\
&&
-6i\pi\left(2 r_H^2+1\right)-24\left[i\pi r_H^4+(1+i\pi)r_H^2+\frac{1}{2}\right]
\ln r_H
\bigg\}.
\end{eqnarray}
Note that the leading contributions of $\mu_{\nu i}^w(Z)$, $\mu_{\nu i}^w(W)$, and $\mu_{\nu i}^w(Z\text{-}H)$ are $\mathcal{O}(m_{\nu i})$, whereas for $\mu_{\nu i}^w(H)$ is $\mathcal{O}(m_{\nu i}^3)$; the latter because $H\nu\bar\nu$ $\propto$ $m_\nu$, which is strongly  suppressed.

\section{Numerical results}

Now we can appreciate the total value of the AWMDM in terms of the neutrino mass  by replacing the corresponding known masses and constants in Eq.~(\ref{AWMDM}):
\begin{eqnarray}\label{AWMDM-numbers}
\mu_{\nu i}^w &=& (5.89+0.28i)\times 10^{-26}\thinspace\frac{m_{\nu i}}{\text{eV}^2},
\end{eqnarray}
where we consider $e=\sqrt{4\pi\alpha(m_Z)}$, $\alpha(m_Z)=1/129$, and the rest of the input values such as the masses of charged leptons, of the gauge bosons, etc. \cite{PDG}. The values of the individual contributing sets are given in Table \ref{TABLE-results}.
As far as individual virtual contribution sets are concerned, the maximum value is due to the $W$ gauge boson one,  with the order of magnitude of $10^{-25}$, however, the contributions with opposite signs coming from the $Z$ and the $Z$-$H$ decrease such maximum value by an order of magnitude, remaining  up to  $10^{-26}$.
A comparison of our results for the  neutrino AWMDM with the well known neutrino AMDM \cite{Fujikawa:1980yx}, given by
\begin{eqnarray}
\mu_{\nu i} &=& \frac{3eG_Fm_{\nu i}}{8\sqrt{2}\pi^2}
=9.49\times 10^{-26}\thinspace\frac{m_{\nu i}}{\textrm{eV}^2},
\end{eqnarray}
yielding
\begin{eqnarray}
\frac{|\mu_{\nu i}^w |}{|\mu_{\nu i}|} &=& 0.62,
\end{eqnarray}
which  implies that the anomalous weak and electromagnetic dipole magnetic  moments are approximately  of the same order of magnitude.

\begin{table}[!t]
  \centering
\begin{tabular}{c|rr}\hline
Contribution &  $\mu_{\nu i}^w$ & \\
\hline
$Z$     & $-(2.58+1.11i)\times 10^{-26}$ & $m_{\nu i}~\textrm{eV}^{-2}$   \\
$W$ & $(1.02+0.14i)\times 10^{-25}$  & $m_{\nu i}~\textrm{eV}^{-2}$   \\
$Z$-$H$ & $-1.69\times 10^{-26}$         & $m_{\nu i}~\textrm{eV}^{-2}$   \\
$H$     & $(1.52+3.13i)\times 10^{-48}$  & $m_{\nu i}^3~\textrm{eV}^{-4}$ \\
Total   & $(5.89+0.28i)\times 10^{-26}$  & $m_{\nu i}~\textrm{eV}^{-2}$   \\
\hline
\end{tabular}
\caption{AWMDM of the Dirac neutrino, the total leading contribution is $\mathcal{O}(m_{\nu i})$, while the pure Higgs contribution is extremely suppressed $\mathcal{O}(m_{\nu i}^3)$.}
\label{TABLE-results}
\end{table}

Now, in order to evaluate the expected numerical values of the neutrino masses in $\mu_{\nu i}^w(m_{\nu i})$, we resort to the neutrino mass analysis given in  Ref.~\cite{Formaggio:2021nfz} from the KATRIN experiment \cite{KATRIN:2019yun} as well as in Ref. \cite{Nu-fit}. Derived from the neutrino oscillations experiments there are two  bounds for the neutrino masses, one scenario is the normal mass-ordering (NMO) and the other one the inverted mass-ordering (IMO).

The NMO, $m_{\nu 1}< m_{\nu 2}< m_{\nu 3}$, arise from the measurements $\Delta m_{21}^2=7.42\times10^{-5}~\text{eV}^2$ and 
$\Delta m_{31}^2=2.517\times10^{-3}~\text{eV}^2$, giving
\begin{equation}\label{}
m_{\nu1}< 1~\text{eV},
\end{equation}
\begin{equation}\label{}
m_{\nu2}=\sqrt{7.42\times10^{-5}~\text{eV}^2+m_{\nu1}^2}
\approx 9\times10^{-3}~\text{eV},
\end{equation}
\begin{equation}\label{}
m_{\nu3}=\sqrt{2.517\times10^{-3}~\text{eV}^2+m_{\nu1}^2}
\approx 5\times10^{-2}~\text{eV},
\end{equation}
from which we can express the AWMDM as a function of the mass of the neutrino $\nu_1$, $\mu_{\nu i}^w(m_{\nu1})$, standing out  that $m_{\nu 1}$ is boundless from below.
The IMO, $m_{\nu 3}< m_{\nu 2}<m_{\nu 1}$, takes form from the measurements $\Delta m_{21}^2=7.42\times10^{-5}~\text{eV}^2$ and 
$\Delta m_{32}^2=-2.498\times10^{-3}~\text{eV}^2$, from which
\begin{equation}\label{}
m_{\nu1}=\sqrt{m_{\nu2}^2-7.42\times10^{-5}~\text{eV}^2},
\end{equation}
\begin{equation}\label{}
m_{\nu2}=\sqrt{2.498\times10^{-3}~\text{eV}^2+m_{\nu3}^2},
\end{equation}
\begin{equation}\label{}
m_{\nu3}< 1~\text{eV},
\end{equation}
so we can express $\mu_{\nu i}^w(m_{\nu3})$, where $m_{\nu 3}$ is boundless from below.

\begin{center}
\begin{figure}[!t]
\subfloat[]{\includegraphics[scale=0.4]{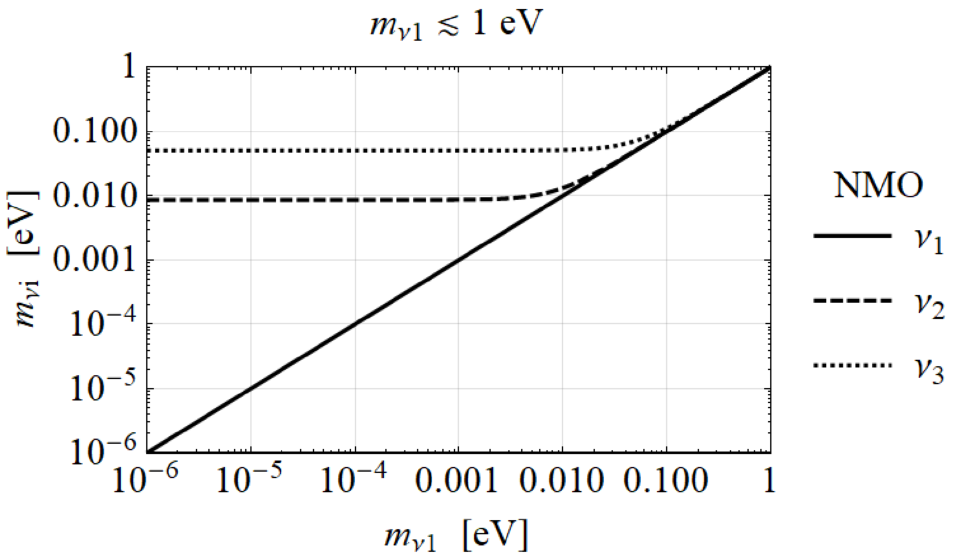}}  \quad
\subfloat[]{\includegraphics[scale=0.4]{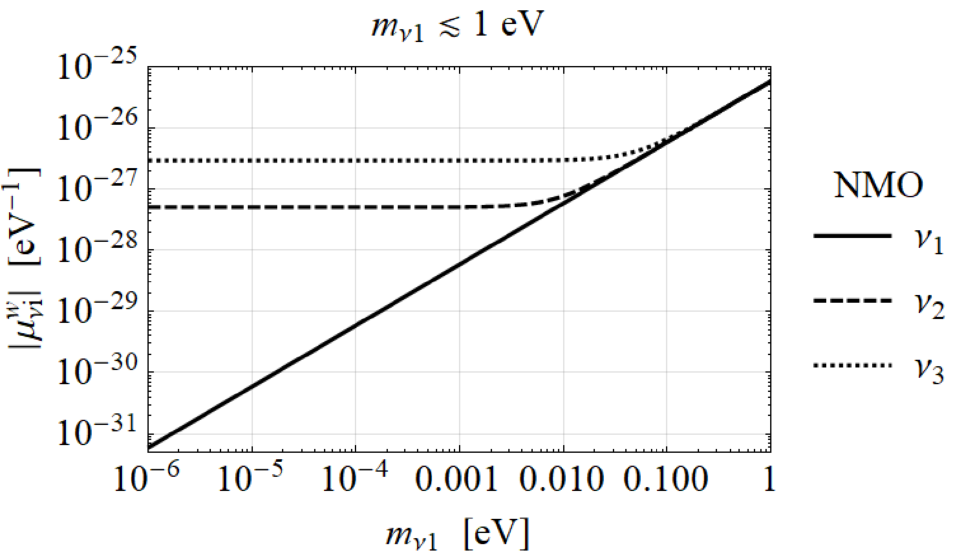}} \\
\subfloat[]{\includegraphics[scale=0.4]{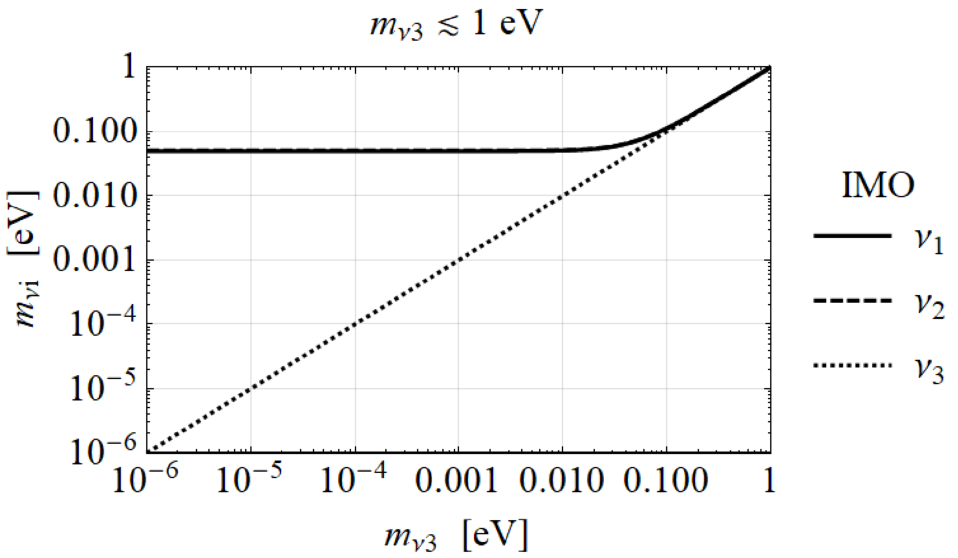}}  \quad
\subfloat[]{\includegraphics[scale=0.4]{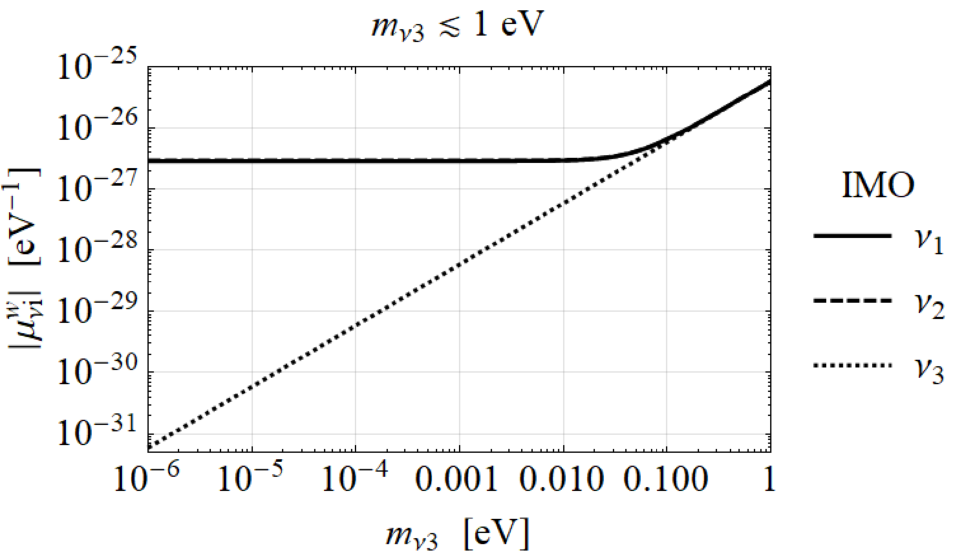}}
\caption{
Normal mass-ordering: (a) neutrino masses with $m_{\nu1}\lesssim 1$ eV, and (b) the magnitude of the AWMDM, which $|\mu_{\nu i}|\lesssim 10^{-26}$~eV$^{-1}$. 
Inverse mass-ordering: (c) neutrino masses with $m_{\nu3}\lesssim 1$ eV, and (d) the magnitude of the AWMDM.}
\label{FIGURE-plotts}
\end{figure}
\end{center}

Therefore, the evaluation of the AWMDM in these two neutrino mass orderings are plotted in Fig.~\ref{FIGURE-plotts}.
The NMO is presented in  Figs.~\ref{FIGURE-plotts}a and \ref{FIGURE-plotts}b: (a) the known mass bounds where   $m_{\nu1}=(10^{-6},1)$ eV, and (b) $|\mu_{\nu i}^w(m_{\nu_1})|$.
The IMO is considered in  Figs.~\ref{FIGURE-plotts}c and \ref{FIGURE-plotts}d: (c) the known mass bounds where $m_{\nu3}=(10^{-6},1)$ eV, and (d) $|\mu_{\nu i}^w(m_{\nu_3})|$. In these plots obviously $|\mu_{\nu i}^w(m_{\nu_1})|$ reproduces the same shape as the neutrino mass plot because it is linearly proportional to the neutrino mass (see Eq.~(\ref{AWMDM-numbers})).

Finally, in order to assess  the impact of the AWMDM in the $Z\to\nu\bar\nu$ decay, we consider the full interaction $Z\nu\bar\nu$, that is to say, the SM vertex plus the dipolar one (MESM). 
Thus, the leading order of $Z\to \nu_i\bar{\nu}_i$, takes the form
\begin{eqnarray}\label{}
\Gamma(Z\to\nu_i\bar{\nu}_i) &\simeq&
\frac{G_Fm_Z^3}{6\sqrt{2}\thinspace\pi}(g_{V\nu_i}^2+g_{A\nu_i}^2)
+\frac{\sqrt{G_F}\thinspace g_{V\nu_i}m_Z^2m_{\nu i}}{\sqrt[4]{128}\thinspace\pi}\left(\mu_{\nu i}^w+\mu_{\nu i}^{w*}\right)
+\frac{m_Z^3}{24\pi}|\mu_{\nu i}^w|^2
\nonumber\\
&=& 1.66\times10^8\thinspace\text{eV}+1.58\times10^{2}\frac{m_{\nu i}^2}{\text{eV}^2}\thinspace\text{eV}+
3.50\times10^{-2}\frac{m_{\nu i}^2}{\text{eV}^2}\thinspace\text{eV},
\end{eqnarray}
where the first term corresponds to the pure SM interaction, the second term is the interference between the SM and the MESM, and the third term is the pure MESM contribution.  Hence, for example, with a neutrino mass of the order of 1 eV, the contribution of the weak dipole moment to the decay process  is   6 orders of magnitude smaller than the resulting from  pure SM.

\section{Conclusions}

We have calculated the weak magnetic dipole moment of Dirac neutrinos in the context of the MESM, where the neutrinos acquire mass by interacting with the Higgs boson, $H\nu\bar\nu\propto m_{\nu}$, as any other SM fermion; such neutrino mass induces one-loop level contributions to the $Z\nu\bar{\nu}$ coupling yielding its AWMDM. In spite that the tree-level SM interaction $Z\nu\bar\nu$ is absolutely dominant, for completeness, we  also analyze its one-loop effect  characterized by the weak dipolar $Z\nu\bar\nu$ coupling. The test of the obtained value of the AWMDM as part of the decay width $Z\to\nu\bar\nu$ shows that the dipole represents a marginal  effect but, nevertheless, a weak property of the neutrino different from zero.

\section*{Acknowledgments}
This work has been partially supported by SNII-Secihti and CIC-UMSNH. JMD thanks to the Secihti program Investigadoras e Investigadores por M\'exico, project 7009. OVH thanks to Secihti for Postdoctoral support.

\end{document}